\documentclass[journal]{vgtc}
\usepackage{listings}
\PassOptionsToPackage{dvipsnames,svgnames}{xcolor}
\usepackage{xcolor} 
\usepackage{graphicx}
\usepackage{tikz}
\usetikzlibrary{overlay-beamer-styles}
\usepackage{subcaption}
\usepackage[most]{tcolorbox}
\usepackage{enumitem}
\usepackage[singlelinecheck=false]{caption}
\usepackage{booktabs}

\definecolor{objectcolor}{HTML}{3D3D3D}
\definecolor{keycolor}{HTML}{A31515}
\definecolor{valuecolor}{HTML}{3D3D3D}
\definecolor{granularbgcolor}{HTML}{B8AAC1} 
\definecolor{intermediatebgcolor}{HTML}{AAB8C1}
\definecolor{highlevelbgcolor}{HTML}{C1B8AA}
\definecolor{systembgcolor}{HTML}{C7D1BC}
\definecolor{commentscolor}{HTML}{008000}
\definecolor{rectanglecolor}{HTML}{b23e3e} 
\definecolor{blockslinecolor}{HTML}{A58070}
\definecolor{edgeslinecolor}{HTML}{708F70}

\newcommand{\schemacomment}[1]{\hspace{0.0em}\textcolor{commentscolor}{\mdseries\sffamily // #1}}

\newtcbox{\systembox}{nobeforeafter, tcbox raise base, rounded corners,
  boxsep=1pt, top=0.2pt, bottom=0.2pt, left=0.2pt, right=0.2pt,
  colback=systembgcolor, colframe=systembgcolor, fontupper={\small\bfseries\sffamily\color{valuecolor}}}

\newtcbox{\highlevelbox}{nobeforeafter, tcbox raise base, rounded corners,
  boxsep=1pt, top=0.2pt, bottom=0.2pt, left=0.2pt, right=0.2pt,
  colback=highlevelbgcolor, colframe=highlevelbgcolor, fontupper={\small\bfseries\sffamily\color{valuecolor}}}

\newtcbox{\intermediatebox}{nobeforeafter, tcbox raise base, rounded corners,
  boxsep=1pt, top=0.2pt, bottom=0.2pt, left=0.2pt, right=0.2pt,
  colback=intermediatebgcolor, colframe=intermediatebgcolor, fontupper={\small\bfseries\sffamily\color{valuecolor}}}

\newtcbox{\granularbox}{nobeforeafter, tcbox raise base, rounded corners,
  boxsep=1pt, top=0.2pt, bottom=0.2pt, left=0.2pt, right=0.2pt,
  colback=granularbgcolor, colframe=granularbgcolor, fontupper={\small\bfseries\sffamily\color{valuecolor}}}

\newcommand{\rectangle}[2]{\tikz[baseline=(Y.base)] \node[draw=none, fill=#1, text=white, rounded corners=2pt, minimum height=1em, inner sep=1pt,align=center,font=\sffamily\scriptsize,anchor=center,text height=1.2ex,text depth=0ex] (Y) {#2};}

\newcommand{\rectanglelegend}[2]{\tikz[baseline=(Y.base)] \node[draw=none, fill=#1, text=black, rounded corners=2pt, minimum height=1em, inner sep=1pt,align=center,font=\sffamily\scriptsize,anchor=center,text height=1.2ex,text depth=0ex] (Y) {#2};}

\NewDocumentCommand{\inlinearrow}{ O{0.8em} O{-0.15em} m }{
  \kern-0.5em
  \raisebox{#2}{
    \includegraphics[height=#1, trim=0 0 0 0, clip]{#3.pdf}
  }
  \kern-0.5em
}

\newcommand{\circl}[2]{\tikz[baseline=-0.05in] \node[draw=none, fill=#1, text=white, circle, draw=black, ultra thin, minimum height=0.5em, inner sep=1pt,align=center,font=\sffamily\scriptsize,anchor=center,text height=1ex,text depth=.25ex] (Y) {#2};}

\onlineid{0}

\vgtccategory{Research}

\vgtcpapertype{please specify}

\title{\projectname: Uncovering Building Blocks\\for Visual Analytics System Design}

\author{%
  Leonardo Ferreira, Gustavo Moreira, and 
  Fabio Miranda
}

\authorfooter{
  \item
  	Leonardo Ferreira, Gustavo Moreira, and Fabio Miranda are with the University of Illinois Chicago.
  	E-mail: \{lferr10, gmorei3, fabiom\}@uic.edu.
}

\abstract{
    Designing and building visual analytics (VA) systems is a complex, iterative process that requires the seamless integration of data processing, analytics capabilities, and visualization techniques.
    While prior research has extensively examined the social and collaborative aspects of VA system authoring, the practical challenges of developing these systems remain underexplored. 
    As a result, despite the growing number of VA systems, there are only a few structured knowledge bases to guide their design and development.
    To tackle this gap, we propose VA-Blueprint, a methodology and knowledge base that systematically reviews and categorizes the fundamental building blocks of urban VA systems, a domain particularly rich and representative due to its intricate data and unique problem sets.
    Applying this methodology to an initial set of 20 systems, we identify and organize their core components into a multi-level structure, forming an initial knowledge base with a structured blueprint for VA system development.
    To scale this effort, we leverage a large language model to automate the extraction of these components for other 81 papers (completing a corpus of 101 papers), assessing its effectiveness in scaling knowledge base construction.
    We evaluate our method through interviews with experts and a quantitative analysis of annotation metrics.
    Our contributions provide a deeper understanding of VA systems' composition and establish a practical foundation to support more structured, reproducible, and efficient system development. VA-Blueprint is available at \href{https://urbantk.org/va-blueprint}{urbantk.org/va-blueprint}.
    
}

\teaser{
  \centering
  \includegraphics[width=\linewidth]{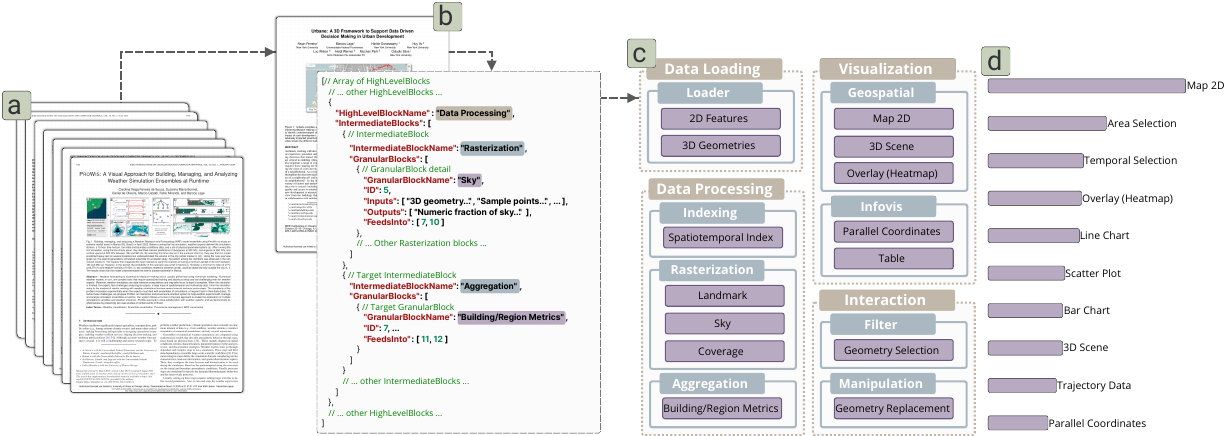}
  \caption{%
    \projectname\ provides a structured way to uncover and organize the fundamental building blocks used in visual analytics (VA) systems.
    (a) The process begins with the curation of a corpus of VA system research papers.
    (b) From each paper, core components and their relationships are systematically extracted and structured into a formal, hierarchical blueprint (represented as JSON), detailing the system's components and their dependencies.
    (c) This blueprint categorizes components across multiple levels of abstraction: high-level stages (e.g., \protect \rectanglelegend{highlevelbgcolor}{Data Processing}), intermediate groups (e.g., \protect \rectanglelegend{intermediatebgcolor}{Rasterization}), and specific granular blocks (e.g., \protect \rectanglelegend{granularbgcolor}{Sky Computation}).
    (d) The final knowledge base comprises 101 machine-readable blueprints encoding component hierarchies, and data and interaction dependencies.
  }
  \label{fig:teaser}
}


\newcommand{\projectname}[0]{VA-Blueprint\xspace}
\newcommand{\myparagraph}[1]{\noindent \textbf{#1.}}
\newcommand{\hide}[1]{}

\newcommand{\highlight}[1]{{#1}}

\graphicspath{{figs/}{figures/}{pictures/}{images/}{./}} 

\usepackage{tabu}                      
\usepackage{booktabs}                  
\usepackage{lipsum}                    
\usepackage{mwe}                       

\usepackage[hyphens]{url}

\usepackage{mathptmx}                  

\usepackage{titlesec}
\titlespacing*{\section}
{0pt} 
{*1.0} 
{*0.4} 

\titlespacing*{\subsection}
{0pt} 
{*0.8} 
{*0.6} 

\titlespacing*{\subsubsection}
{0pt} 
{*0.8} 
{*0.6} 

\begin{document}

\maketitle


\section{Introduction}

Visual analytics (VA) systems help users make sense of complex data by combining analytics with interactive visualizations.
Their usefulness has been demonstrated across diverse domains, including biology~\cite{krueger_facetto_2020}, healthcare~\cite{kwon_retainvis_2019}, urban planning~\cite{ferreira_urbane_2015}, and climate science~\cite{gautier_co-visualization_2020}.
However, building VA systems remains a highly complex and iterative process.
System builders must integrate data processing techniques, analytics capabilities, and visualization strategies while ensuring interpretability for domain experts and maintaining efficiency to support interactivity.
Despite their significance and inherent challenges, VA system development is often approached in an ad-hoc manner, with minimal reuse of existing components. Given the necessity to address the unique analytical and visualization needs of specific domains, these systems are usually bespoke solutions, leading to fragmented development efforts, making them difficult to extend, adapt, or reuse across projects.
This tension, between the need for tailored solutions and the demand for scalable, extensible systems, presents a fundamental challenge in VA system development~\cite{cui_visual_2019, wu_grand_2023}. On one hand, VA systems must be grounded by domain-specific data, tasks, and analytical workflows: a visualization technique that works well for transportation may be inadequate for urban planning, just as an analytical model for climate science may not be applicable to public health.
On the other hand, as VA systems become increasingly complex, the lack of systematic and reusable components hampers the design process and increases the effort needed for development.

Currently, there is a lack of practical knowledge bases that document the building blocks of VA systems.
Initial steps have produced valuable taxonomies for visualization tasks and techniques \cite{brehmer_multi-level_2013}. 
More recently, new initiatives have emerged to catalog and structure design elements derived from system surveys~\cite{ying_vaid_2024}.
However, these contributions often focus on specific aspects (primarily visualization) or lack the granular, interconnected structure needed to capture the full dataflow and component composition of entire VA systems.
Consequently, a practical, component-level blueprint encompassing data processing, analytics, visualization, and interaction is still missing.
Without such a foundation, developers are forced to either create highly specialized systems from scratch or repurpose existing solutions that may not fully meet their needs.
The status quo is a landscape where many VA systems remain one-off, non-scalable, and difficult to extend, limiting broader adoption and cross-domain innovation.
As VA systems become more widespread in different domains and increasingly applied in decision-making scenarios~\cite{urban_weil_2023}, there is a growing need to bridge this gap.
Fundamentally, \emph{how can we design VA systems that remain useful for specific datasets, tasks, and users while also being extensible and reusable?}

In this paper, we take a step towards answering this question by proposing a structured methodology to identify, categorize, and organize VA system components.
Given the broad and diverse applications of VA, we focus our efforts on urban VA systems. These systems span multiple domains, including transportation, urban planning, and environmental science.
Through a systematic process, we analyze existing urban VA systems, extracting and classifying their core building blocks.
Using an initial sample of 20 systems, we create \textbf{VA-Blueprint} (Figure \ref{fig:teaser}), a multi-level knowledge base that provides a structured representation of these components.
Then, to scale this effort, we investigate the potential of a large language model (LLM) to automate the extraction of VA system components from a corpus of 101 research papers, building on demonstrated LLM capabilities for related tasks such as aspect-based summarization \cite{yang_exploring_2023}.
We assess LLM's ability to identify and categorize these components.
Through this approach, we move towards addressing the challenge of VA system development with a structured methodology that balances the need for bespoke, domain-specific solutions with the benefits of extensible and reusable systems.
\highlight{Concretely, our work offers a data-driven catalog of urban VA patterns that supports VA system development in three key ways: (1) enabling pattern reuse, allowing developers to identify and draw inspiration from proven configurations instead of assembling ad-hoc pipelines; (2) laying the foundation for model-driven development~\cite{selic_pragmatics_2003} in urban VA; and (3) ultimately guiding developers in building more effective urban VA systems by highlighting common, effective architectural structures.}
%
%
The knowledge base and associated tools are publicly available at \href{https://urbantk.org/va-blueprint}{urbantk.org/va-blueprint}.

\section{Related Work}

In this section, we review prior work in two key areas of relevance to this paper.
First, we review methodologies that support the design of VA systems, as well as how system components, design patterns, and best practices are documented and shared.
Then, we review prior work on the use of LLMs for extracting information from literature.

\subsection{Methodologies \& knowledge sharing for VA systems}

The design of VA systems is deeply rooted in a close collaboration between visualization researchers, practitioners, and domain experts.
While the nature of this collaboration varies in structure and intensity~\cite{kirby_visualization_2013}, it usually involves visualization researchers contributing technical expertise to elicit system requirements and develop visualizations, interactions, and analytical components, while domain experts provide problem definitions, tasks, and data.
This process is often guided by human-centered methodologies that structure the design workflow~\cite{wu_grand_2023}, offering guidelines for key stages such as problem characterization and visualization development~\cite{sedlmair_design_2012, wu_grand_2023}.
However, recent studies have shed light on the tensions within this process, both from a social and a technological perspective.
From a social perspective, Akbaba et al.~\cite{akbaba_troubling_2023} discussed challenges in visualization collaborations, emphasizing the need for stakeholders to recognize benefits beyond the tool itself.
Wu et al.~\cite{wu_defence_2022} reported a series of criticisms aimed at VA systems, including growing concerns regarding the generalizability of contributions, a concern echoed by several works~\cite{ weber_apply_2017, meyer_criteria_2020, wu_grand_2023}.
From a technological perspective, Isenberg recently highlighted reproducibility challenges in visualization research~\cite{isenberg_state_2024}.
Chen and Ebert discussed the many challenges of designing VA systems, pointing to the trial-and-error nature of VA system design~\cite{chen_ontological_2019}, an often iterative and unpredictable process.
At the intersection of these concerns, Wu et al.~\cite{wu_defence_2022} argued for the augmentation of traditional HCI-grounded design study methodologies with software engineering perspectives to enable more systematic development and evaluation of VA systems.
In summary, prior research has shown that VA system design is a complex and iterative endeavor, often relying on the experience and intuition of researchers~\cite{chen_ontological_2019}.

To mitigate these challenges and lower the barriers to design and development, several studies have proposed toolkits and knowledge bases to support the structured creation of these systems.
At their core, these efforts aim either to operationalize design spaces (streamlining system construction from a practical perspective) or to formalize taxonomies of components from a theoretical perspective~\cite{chen_pathways_2017}.
These goals align with the broader challenge of moving away from monolithic systems towards more modular ones~\cite{wu_grand_2023}.
From a practical perspective, toolkits and frameworks have played a key role in facilitating design by providing components that simplify system development and enable rapid prototyping of ideas, while reducing the need for low-level coding~\cite{mcnuut_grammar_2023, ferreira_assessing_2024}.
Vega-Lite~\cite{satyanarayan_vega-lite_2017} and Draco~\cite{moritz_formalizing_2019} exemplify how declarative specifications can simplify visualization construction and embed best practices.
Similar toolkits and frameworks have been proposed taking into account specific contexts and domains, such as virtual reality~\cite{sicat_dxr_2019}, uncertainty visualization~\cite{kay_ggdist_2024}, urban analytics~\cite{moreira_urban_2024}, and genomics~\cite{gosling_lyi_2022}.
From a theoretical perspective, taxonomies offer a vocabulary that can bridge communication gaps and reduce misunderstandings~\cite{chen_pathways_2017}. While taxonomies have been widely adopted in visualization research in general~\cite{duke_building_2004, duke_do_2005, gilson_from_2008, perez_enhanced_2011, polowinski_viso_2013}, relatively few attempts have been made to create taxonomies specifically for VA systems (e.g.,~\cite{sacha_vis4ml_2019, chen_ontological_2019}).
A notable example is the work by Ying et al.~\cite{ying_vaid_2024}, where they presented a knowledge base of VA views, based on a survey of 124 systems. The knowledge base is made available through an indexing scheme composed of task and design, based on the multi-level typology of visualization tasks~\cite{brehmer_multi-level_2013}, facilitating the ideation process through the exploration of previous VA designs.
However, despite their potential to enhance knowledge sharing, reproducibility, and modularity, taxonomies in VA and their practical software-oriented implementations remain underdeveloped as they rely on manual curation and expert-driven classification. Constructing and maintaining these taxonomies requires substantial effort.

Our work moves beyond these approaches by exploring a semi-automated, bottom-up method for extracting VA system components directly from research papers.
Given that many systems are not publicly available, this approach enables systematic identification and classification of components based solely on the literature.





\subsection{Automated approaches for knowledge extraction}

The visualization community has long explored automated ways to extract and organize information from scientific literature and repositories, given the labor-intensive process of manually curating knowledge bases.
Li et al.~\cite{li_structure-aware_2022} proposed a method to retrieve visualizations taking into account their perceptual similarities.
Poco and Heer~\cite{poco_reverse-engineering_2017} leveraged machine learning to extract visual encoding specifications from images.
Hoque and Agrawala~\cite{hoque_searching_2020} presented a search engine based on over 7,000 D3 visualizations crawled from the web.
Li et al.~\cite{li_utilizing_2021} introduced a novel image-based representation designed to encode information from biomedical papers and support more effective indexing.
Other works have proposed extracting information and insights from static charts to support the generation of animated visualizations~\cite{ying_reviving_2024}, or to derive insights and descriptive text~\cite{zhou_intelligent_2023}.

Concurrently, LLMs have been shown to effectively extract information from textual data, enabling applications such as data summarization and report generation~\cite{minaee_large_2025}.
Zhang et al.~\cite{zhang_systematic_2024} and Pu et al.~\cite{pu_summarization_2023} highlighted the disruptive change brought forward by LLMs for summarization and their capabilities even for zero-shot summarization.
A similar conclusion is reached by Agarwal et al.~\cite{agarwal_litllms_2025} but for the automatic generation of literature reviews.
In the visualization community, Tang et al.~\cite{tang_steering_2024} proposed to use intermediate workspaces to steer the summarization of documents, including literature reviews.
Despite these advancements, most existing approaches have concentrated on isolated chart analysis or textual summarization.
In contrast, the systematic extraction of VA system components, particularly using LLMs, remains an underexplored direction.
%
%
Rather than focusing solely on retrieving visualization elements or summarizing papers, our approach seeks to identify the analytical, interactive, and visualization components that underpin VA systems.
By systematically analyzing research literature, we construct a knowledge base that captures the modular building blocks of these systems.

\section{Building a Knowledge Base for VA Systems}
\label{sec:methodology}

Building VA systems is a complex process that involves integrating data processing, analytical techniques, and interactive visualizations.
However, despite the growing number of proposed systems, their underlying components remain poorly documented and dispersed across the literature.
In the absence of a structured approach to \emph{capture} and \emph{organize} this knowledge, designing new VA systems often requires starting from scratch or relying on researchers' intuition.
To tackle this gap, we propose \projectname.
\projectname\ is a knowledge base composed of structured descriptions of VA systems.
\projectname\ can be framed from three different perspectives, each highlighting a distinct aspect of our contributions. Together, these perspectives help articulate the core research questions that guide our study.

\myparagraph{[1] The need for systematic knowledge extraction}
There is no structured repository that captures the fundamental components of VA systems, despite their increasing complexity.
This increases over-reliance on researchers' experience.
And while taxonomies exist for specific visualization techniques, \emph{how can we systematically extract and categorize VA system components from research papers?}

\myparagraph{[2] The scalability challenges in VA knowledge organization}
Traditional taxonomy creation relies on researchers' expertise and manual effort, which does not scale as new systems emerge.
Then, \emph{how effective are automated methods in identifying VA system components, and what are their limitations in this context?}

\myparagraph{[3] Validating extracted VA components}
Equally important to extracting VA components is ensuring that they align with expert knowledge and can be validated for accuracy.
So \emph{what validation methods can be used to assess accuracy?}

Next, we detail our methodology, which is guided by these research questions.
Specifically, we describe: the requirements that \projectname\ must fulfill (Section~\ref{sec:requirements}); our methodology (Section~\ref{sec:methodologyoverview}); the approach for curating a corpus of VA system papers (Section~\ref{sec:corpus}); the process of structuring VA systems (Section~\ref{sec:manual}); and the use of LLMs to extract components (Section~\ref{sec:llm}).

\subsection{Knowledge base requirements}
\label{sec:requirements}

Our knowledge base must fulfill several key requirements.
These requirements were derived from our review of the literature, two prior surveys on domain-specific VA systems~\cite{miranda_state_2024, ferreira_assessing_2024}, and our experience as system builders.
Given that research papers are often the only concrete artifacts documenting VA system designs, our knowledge base must be structured in a way that enables the extraction and organization of components from them.
We define the following requirements:

\myparagraph{[R1] Granularity of components}
The knowledge base must represent VA system components at multiple levels of abstraction, distinguishing between high-level functional categories (e.g., data processing, analytics, visualization) and low-level implementations (e.g., specific algorithms, techniques, libraries).

\myparagraph{[R2] Categorization of components}
The extracted components must be organized based on their role within a VA system. For example, components must be categorized with respect to their role in data operation, analytics, interaction, and visualization.

\myparagraph{[R3] Connection between components}
Within a VA system, components are interconnected, not isolated.
The knowledge base must then capture their dependencies and relationships.

\myparagraph{[R4] Contextual information}
Each component should include metadata containing detailed information such as its definition, usage scenarios (when available), source references, and related components.

\myparagraph{[R5] Queryable structure}
Entries in the knowledge base must be queryable, enabling users to explore system components based on different criteria (e.g., component type, functionality).

\subsection{Methodology overview}
\label{sec:methodologyoverview}

\begin{figure}
  \centering
  \includegraphics[width=1\linewidth]{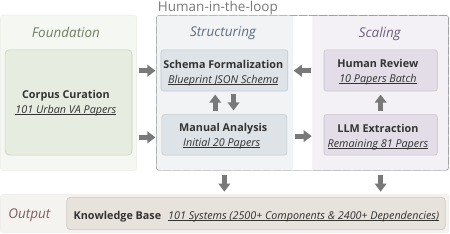}
  \caption{%
  VA-Blueprint knowledge base construction process overview.
  \textbf{\textit{Foundation}} comprises curating the paper corpus (101 papers).
  In \textbf{\textit{Structuring}}, manual analysis of an initial set of papers (20) establishes core components and informs the formal JSON schema.
  \textbf{\textit{Scaling}} uses LLM extraction for the remaining papers (81), guided by the schema and refined in a human-in-the-loop review cycle.
  The result is the final knowledge base (\textbf{\textit{Output}}) documenting 101 systems.
  }
  \label{fig:methodology}
\end{figure}

\highlight{Our methodology comprises a set of stages, as illustrated in Figure~\ref{fig:methodology}.}
\highlight{In the \textbf{\textit{Foundation}} stage, we collect a set of VA papers (Section~\ref{sec:corpus}).}
\highlight{This is followed by an iterative, human-in-the-loop process to construct the knowledge base.}
\highlight{This process cycles between: (1) a \textbf{\textit{Structuring}} stage, where we manually analyze an initial set of papers to establish a formal schema and a core set of components (Section~\ref{sec:manual}), and (2) a \textbf{\textit{Scaling}} stage, where we leverage an LLM to automate the extraction for the remaining corpus (Section~\ref{sec:llm}).}
%
%
%
%

\subsection{Foundation: Corpus curation}\label{sec:corpus}
%
%
%
In this paper, we focus on a specific area of VA: systems designed for urban analyses. This decision is motivated by two factors.
First, urban VA systems are highly diverse in terms of data sources and system components, requiring the integration of data processing, analytics, and visualizations to handle often large, heterogeneous, and multi-scale datasets. The complexity of these systems makes them representative examples of VA architectures.
Second, urban VA serves a broad range of users and domain experts, including urban planners, transportation specialists, engineers, public health experts, and climate scientists. These systems must then accommodate diverse workflows, making them a compelling domain for studying how VA components are structured and interconnected.
Such delineation of our corpus allows us to extract a diverse set of VA systems while maintaining a cohesive scope that facilitates meaningful comparisons and categorization.

We establish the following criteria for paper selection: (1) papers must describe a VA system that integrates data processing, analytics, and visualization components, rather than focusing solely on visualization techniques; (2) the system must be applied to an urban analytics problem; (3) the paper must have been published in a top-tier visualization venue (e.g., IEEE VIS, EuroVis, PacificVis, IEEE TVCG, CGF, C\&G).
For an initial selection of papers, we leveraged our previous surveys on urban VA~\cite{miranda_state_2024, ferreira_assessing_2024}, which provided an initial pool of over 85 papers.
These prior surveys examined the landscape of urban VA systems.
From this set, we manually reviewed papers to further refine our corpus, ending with 101 papers.
We tried to balance the papers based on their areas of application (e.g., transportation, planning).
This set of 101 papers will serve as the basis for our extraction methodology.
Figure~\ref{fig:methodology} shows an overview of the knowledge base construction process.
Our corpus spans multiple areas (e.g., transportation, climate,  accessibility), under the umbrella of urban analytics.
The research labs contributing to this corpus are also well distributed, with contributions from institutions based in the Americas, Europe, and Asia.

\begin{tcolorbox}[
    colback=gray!3,
    colframe=gray!30,
    arc=1pt,
    title={VA-Blueprint Schema},
    coltitle=black,
    size=small
    ]
\begin{description}[leftmargin=0cm, labelsep=0.25em, itemsep=0.5em,
                    style=sameline, font={\small\bfseries\sffamily},
                    before={\small\sffamily}]

    \item[\systembox{SystemBlueprint} \{] ~\\
        \hspace*{1em} \textcolor{keycolor}{PaperTitle}: \textcolor{valuecolor}{string;} \schemacomment{Paper Title} \\
        \hspace*{1em} \textcolor{keycolor}{HighLevelBlocks}: \textcolor{valuecolor}{HighLevelBlock[];} \schemacomment{List of high-level blocks} \\
        \hspace*{0em} \}

    \item[\highlevelbox{HighLevelBlock} \{] ~\\
        \hspace*{1em} \textcolor{keycolor}{HighLevelBlockName}: \textcolor{valuecolor}{string;} \schemacomment{High-level block's name} \\
        \hspace*{1em} \textcolor{keycolor}{IntermediateBlocks}: \textcolor{valuecolor}{IntermediateBlock[];} \schemacomment{List of intermediate blocks} \\
        \hspace*{0em} \}

    \item[\intermediatebox{IntermediateBlock} \{] ~\\
        \hspace*{1em} \textcolor{keycolor}{IntermediateBlockName}: \textcolor{valuecolor}{string;} \schemacomment{Intermediate block's name} \\
        \hspace*{1em} \textcolor{keycolor}{GranularBlocks}: \textcolor{valuecolor}{GranularBlock[];} \schemacomment{List of granular blocks} \\
        \hspace*{0em} \}

    \item[\granularbox{GranularBlock} \{] ~\\
        \hspace*{1em} \textcolor{keycolor}{GranularBlockName}: \textcolor{valuecolor}{string;} \schemacomment{Granular block's name} \\
        \hspace*{1em} \textcolor{keycolor}{ID}: \textcolor{valuecolor}{integer;} \schemacomment{Unique block identifier} \\
        \hspace*{1em} \textcolor{keycolor}{PaperDescription}: \textcolor{valuecolor}{string;} \schemacomment{Component's summarized description} \\
        \hspace*{1em} \textcolor{keycolor}{Inputs}: \textcolor{valuecolor}{string[];} \schemacomment{Data/signals consumed by the block} \\
        \hspace*{1em} \textcolor{keycolor}{Outputs}: \textcolor{valuecolor}{string[];} \schemacomment{Data/signals produced by the block} \\
        \hspace*{1em} \textcolor{keycolor}{ReferenceCitation}: \textcolor{valuecolor}{string;} \schemacomment{Quote from the paper} \\
        \hspace*{1em} \textcolor{keycolor}{FeedsInto}: \textcolor{valuecolor}{integer[];} \schemacomment{IDs of downstream blocks} \\
        \hspace*{0em} \}

\end{description}
\end{tcolorbox}

\subsection{Structuring: Schema and manual analysis}
\label{sec:manual}
Building on the foundational corpus and blueprint (Section~\ref{sec:corpus}), the \textbf{\textit{Structuring}} stage translates the conceptual model into a practical, scalable format. 
\highlight{This stage involved (1) developing the formal multi-level representation and its JSON encoding, and (2) performing manual analysis of selected papers to instantiate and validate this structure.}



\subsubsection{VA specification schema}

%
\highlight{
We model VA systems as multi-level dataflows composed of interconnected components and operations, with explicit data and interaction dependencies. This formal model both guides our understanding of VA system architectures and defines the structure of our specification schema, which is used for automated extraction and validation.
}
%
%
%
As highlighted in previous works~\cite{sedlmair_design_2012}, VA systems generally consist of three core building blocks: data processing, analytics, and visualization.
These systems operate as layered computational workflows, where data undergoes a series of processing and analytics steps until visualization.
While VA systems can be described using UML diagrams, architecture diagrams, or formal specifications, we adopt a dataflow representation as it more naturally captures how data is transformed and propagated across system components~\cite{nonnemann_characterization_2020}.
Most VA systems, even if they do not explicitly expose a dataflow to users~\cite{yu_visflow_2017, chen_vaud_2018, moreira_curio_2025}, implicitly follow this structure through function calls, API interactions, or event-driven communication between components.
The concept of data being transformed through a series of operators was also discussed as the \emph{operator} pattern in Heer and Agrawala's design patterns for visualization software~\cite{heer_software_2006}.
This pattern has been extensively used for intra-tool communication, reinforcing its relevance as a guiding model for our extraction of VA system components~\cite{nonnemann_characterization_2020}.

Referring back to our goal of incorporating multiple levels of abstraction, we model VA system components across three hierarchical levels: system-level, component-level, and operation-level representations.
\highlight{This model directly defines the structure of our machine-readable specification schema, implemented as a hierarchical JSON format.}
The specification schema serves three roles: (1) structuring the LLM’s output by providing a template for extraction, (2) ensuring alignment with our formal model, and (3) enabling parsing and validation for downstream analysis and visualization.
The VA-Blueprint Schema illustrates the core hierarchical structure and key fields defined in this specification.
As a running example, we will use the description of an urban VA system composed of a map, scatter plot, and bar chart that visualizes topological features extracted from spatiotemporal data.



\noindent \protect\circl{systembgcolor}{} \textbf{System-level representation.}
At the highest level, we define that a VA system can be represented as a directed graph $\mathcal{S} = (\mathcal{C}, \mathcal{D})$, where $\mathcal{C}$ is the set of components of the system, and $\mathcal{D}$ represents dependencies between components.
Each dependency in $\mathcal{D}$ represents either a data dependency (i.e., data transfer between components or operations) or an interaction dependency (i.e., constraints or filters imposed by user interaction).
The schema encodes this structure in the \rectanglelegend{systembgcolor}{SystemBlueprint}, listing components and their interconnections.

\noindent\emph{Example.}
In our example system, we model a VA system that includes a spatial data processing component $C_{spatial}$, a topological feature extraction component $C_{topo}$, and three visualization components: $C_{map}$, $C_{scatter}$, and $C_{bar}$. These components interact through structured dependencies, where data processing feeds into analysis and visualization components.

\noindent \protect\circl{highlevelbgcolor}{}~\protect\circl{intermediatebgcolor}{} \textbf{Component-level representation.}
A VA component is defined as a unit $ C = (\mathcal{I}, \mathcal{P}, \mathcal{O}, \mathcal{T})$ with inputs $\mathcal{I}$, properties $\mathcal{P}$, outputs $\mathcal{O}$, and internal operations $\mathcal{T}$.
These map to \rectanglelegend{highlevelbgcolor}{HighLevelBlocks} and \rectanglelegend{intermediatebgcolor}{IntermediateBlocks} in the specification, describing structure and functional roles.

\noindent\emph{Example.}
In our example system, $C_{spatial}$ includes operations for data binning ($T_{bin}$) and density computation ($T_{density}$). The topological feature extraction component $C_{topo}$ depends on the output of $T_{density}$ to compute topological features. Visualization components ($C_{map}$, $C_{scatter}$, $C_{bar}$) then receive processed data to create representations for the user.

\noindent \protect\circl{granularbgcolor}{} \textbf{Operation-level representation.}
At the lowest level, an operation is an atomic function $T_i: I_i \times P_i \to O_i$, capturing low-level tasks within components.
Each operation $T_i$ is a function within a component that executes a specific task.
Each operation $T_i$ belongs to at least one component but may be shared across multiple components.
These operations map to \rectanglelegend{granularbgcolor}{GranularBlocks}, detailing inputs, outputs, and downstream links.

\noindent\emph{Example.}
For instance, $T_{density}$ is used in both $C_{spatial}$ and $C_{map}$, while $T_{bin}$ is necessary for both spatial data processing and the bar chart aggregation ($C_{bar}$). The operations $T_{topo}$ and $T_{feature}$ are specific to $C_{topo}$ for extracting structural features from spatial datasets.

\noindent\textbf{Dependencies between components and operations.}
A component may depend on one or more other components, meaning it receives inputs from multiple sources. We denote this as $\{C_{j_1}, C_{j_2}, ..., C_{j_k}\} \rightarrow C_i$, indicating that $C_i$ requires the outputs of $C_{j_1}, C_{j_2}, ..., C_{j_k}$.
Similarly, operations can depend on the outputs of other operations: $\{T_{j_1}, T_{j_2}, ..., T_{j_k}\} \rightarrow T_i$.
Since operations may belong to multiple components, we also express $\{T_{j_1}, T_{j_2}, ..., T_{j_k}\} \rightarrow C_i$, meaning the execution of these operations produces outputs required by $C_i$.
%
%
We classify dependencies into two types: (1) Data dependencies, which describe how data flows through the system -- where the output of one component or operation serves as input to another; and (2) Interaction dependencies, which capture how user-driven interactions affect downstream components or operations, for example by passing constraints or filters rather than raw data.
Data dependencies will be created when a component or operation produces data that is used as input for another component or operation. Interaction dependencies will be created when a component constrains or modifies another component or operation.

\noindent\emph{Example.}
In our example, density transformation provides data to the topology component: $(T_{density}, C_{topo})$.
In turn, the scatter plot depends on processed features from the topology component: $(C_{topo}, C_{scatter})$.
The bar chart requires time-aggregated data from the binning transformation: $(T_{bin}, C_{bar})$.
For interaction dependencies, the scatter plot supports user-driven filtering in both the map and the bar chart: $\{(C_{scatter}, C_{map}), (C_{scatter}, C_{bar})\}$.

\subsubsection{Initial knowledge base}
\label{sec:manual_analysis}
%
\highlight{To align the specification schema with real-world systems, we manually analyzed 20 research papers from our corpus, constructing an initial knowledge base of VA system components, operations, and dependencies.}
\highlight{This initial set was deliberately curated to ensure diversity across different dimensions.}
\highlight{The selection included systems from various application domains (e.g., transportation~\cite{weng_towards_2021}, urban planning~\cite{lyu_if-city_2024}, flooding ~\cite{boorboor_submerse_2024}, weather~\cite{de_souza_prowis_2022}), using a wide range of data types (e.g., POI~\cite{chen_sensemap_2024}, crime~\cite{garcia_crimanalyzer_2021}, social media~\cite{cao_whisper_2012}), and originating from different authors and research groups.}
%
Through a systematic review, we identified system components, operations, and dependencies.
\highlight{We selected 20 papers because, during the review process, we observed that the discovery of new component labels reached saturation. Specifically, the relative rate of new label discovery fell below 3\% by the 20th paper, indicating that additional papers were unlikely to contribute substantially novel components.}

In this review, we considered a component to be a self-contained, identifiable computational unit within the system. Conceptually, it should be modular enough to be reused and essential enough that removing it would degrade the system's functionality or structure.
When defining system components, we aimed to straddle a balance between over-generalization (i.e., they are too abstract to be useful in dataflows) and over-specialization (i.e., they are too rigid and unique per system, preventing reuse and modularity).
To achieve this balance, we established a set of guiding principles for identifying and structuring components.
First, components should encapsulate well-defined functionality.
Second, they should have clear input-output relationships.
Third, components should be generalizable to units that can be used in different systems.
Fourth, if the absence of a particular component breaks the dataflow, it is a core component.
During our review, we categorized components into data processing (e.g., data ingestion, transformation, filtering), analytics (e.g., clustering algorithms, feature extraction), visualization units, and interaction types.
We analyzed both the textual descriptions and figures, focusing on sections detailing system architecture and implementation.

In addition, we examined operations, the low-level computations performed within a component. These serve as the building blocks of each component, specifying how input data is processed, transformed, or rendered.
We map operations to components, establishing a hierarchical representation of VA systems.
Finally, we also documented dependencies between components and operations.
Textual descriptions usually provided clear cues regarding dependencies.
For interaction dependencies, we paid attention to mentions of \emph{linked brushing}, \emph{linked views}, or \emph{filtering}.

For each of the 20 selected papers, we conducted an in-depth review to identify an initial set of components, operations, and dependencies through an open coding approach.
The process was carried out by a team of three researchers (one faculty member and two graduate assistants).
Each paper was reviewed by one graduate assistant, with weekly meetings to review, refine, and resolve disagreements through discussion between all team members.
In this process, we surfaced shared elements and developed a vocabulary of elements across systems.
All identified codes were collected into a shared pool of components, operations, and dependencies.
Once the initial set of codes was compiled, we grouped conceptually similar elements and refined their labels for naming consistency and functional clarity. For example, different terms for clustering (e.g., ``clustering'', ``grouping'') were unified under the cluster operation label and interaction techniques such as ``brushing'' or ``highlighting'' were grouped under select.
After coding all 20 papers, we performed a final pass, making sure that the elements were aligned with our formal model.
The final set of elements was reviewed and approved by all team members.

\subsection{Scaling: Using an LLM to extract VA knowledge}
\label{sec:llm}

While manually analyzing 20 papers allowed us to construct a consistent initial knowledge base, this process is time-consuming and difficult to scale. On average, it took approximately one hour to thoroughly review, code, and categorize each paper.
To scale the knowledge base, we propose leveraging LLMs to automatically extract system elements from research papers.
Our approach builds on the formal representation and codebook developed in the manual phase.

\subsubsection{Prompting the LLM with extraction tasks}
\label{sec:llm_extraction}

Once the schema was defined, we developed a prompting approach to guide an LLM in extracting system specifications from research papers. 
We used OpenAI's GPT-4.
Modern LLMs like GPT-4 demonstrate strong capabilities in zero-shot or few-shot learning for structured information extraction tasks, particularly when guided by well-defined schemas \cite{caufield_structured_2024}. 
The detail and hierarchical structure of our JSON schema are designed to leverage this strength, providing explicit constraints and context that help the model accurately map textual descriptions to the target data structure.
To further enhance reliability and ensure the output precisely matched our desired format, we employ few-shot prompting with three specifications from our initial knowledge base to show the model what an ideal response looked like.
Each prompt was designed to guide the LLM through a multi-step extraction process.
At its core, the prompt included:

\begin{itemize}[nosep, noitemsep,topsep=0pt, leftmargin=*]
    \item A description of the task. The model was told to act as a system designer reviewing a research paper to extract a complete specification of a VA system.
    \item A detailed explanation of the schema. We included the full JSON schema as a reference.
    \item Instructions for document coverage. We asked the model to read the entire paper beyond the abstract and system overview, also including methods, implementation details, and use cases.
    \item Few-shot examples. We included examples from the manually created set.
\end{itemize}

To ensure robustness, we incorporated standard practices from LLM summarization. 
\highlight{This was implemented through a human-in-the-loop process (Figure~\ref{fig:methodology}), where we engaged in cycles of review and refinement.}
\highlight{As part of this process, if an output was incomplete, ambiguous, or inconsistent upon review, we created follow-up prompts asking the model to revise specific sections based on previous results.}
\highlight{Our manual refinement involved relabeling $\sim$10\% of blocks for terminological consistency and adding $\sim$5\% of missing ones; erroneous block removals were rare.}
\highlight{The more substantial effort was adjusting $\sim$30\% of dependency edges, which mostly involved adding overlooked connections to complete the system dataflow.}
%
In the last step, once we agreed that the model returned a correct JSON specification, we parsed and checked the output for schema validity.

\section{\projectname: A knowledge base for VA systems}
Following the methodology detailed in Section \ref{sec:methodology}, we constructed the \projectname\ knowledge base—a structured repository capturing the constituent components of urban VA systems.
This section first provides an overview of the knowledge base's structure and content (Section \ref{sec:kboverview}).
We then introduce the visual interface developed to navigate and validate its entries and discuss the accessibility of these resources (Section \ref{sec:interface}).

\subsection{Knowledge base overview}
\label{sec:kboverview}
The \projectname\ knowledge base currently comprises structured representations of 101 urban VA systems, extracted systematically from research literature.
Each system entry is stored as a distinct JSON file, adhering to the multi-level blueprint explained in Section~\ref{sec:llm}.
This structure organizes components hierarchically:
\begin{enumerate}[noitemsep, topsep=0pt, leftmargin=*]
    \item \textbf{High-level Blocks:} Represent the major stages in a VA system: \rectanglelegend{highlevelbgcolor}{Data Loading}, \rectanglelegend{highlevelbgcolor}{Data Processing}, \rectanglelegend{highlevelbgcolor}{Visualization}, and \rectanglelegend{highlevelbgcolor}{Interaction}. These align with the primary phases of a typical VA workflow.

    \item \textbf{Intermediate Blocks:} Provide a more specific functional grouping within the high-level blocks. Common examples derived during our curation include \rectanglelegend{intermediatebgcolor}{Loader} (under \rectanglelegend{highlevelbgcolor}{Data Loading}); \rectanglelegend{intermediatebgcolor}{Querying}, and \rectanglelegend{intermediatebgcolor}{Clustering} (under \rectanglelegend{highlevelbgcolor}{Data Processing}); \rectanglelegend{intermediatebgcolor}{Geospatial} and \rectanglelegend{intermediatebgcolor}{Infovis} (under \rectanglelegend{highlevelbgcolor}{Visualization}); and \rectanglelegend{intermediatebgcolor}{Filter} and \rectanglelegend{intermediatebgcolor}{Annotation} (under \rectanglelegend{highlevelbgcolor}{Interaction}). These intermediate blocks capture recurring sub-tasks or component types within the broader categories.

    \item \textbf{Granular Blocks:} Represent the most specific elements identified in each system. These granular components include concrete visualization types (e.g., \rectanglelegend{granularbgcolor}{Map 2D}, \rectanglelegend{granularbgcolor}{Line Chart}), specific interaction mechanisms (e.g., \rectanglelegend{granularbgcolor}{Area Selection}), distinct data inputs (e.g., \rectanglelegend{granularbgcolor}{Trajectory Data}), or particular analytical methods (e.g., \rectanglelegend{granularbgcolor}{k-Means Clustering}). Each granular block includes details about its inputs, outputs, description, and an exact reference from the source paper that verifies the block’s existence.

    \item \textbf{Dependencies:} Edges capture the relationships between blocks at various levels, categorized as either data dependencies (representing data flow) or interaction dependencies (representing control flow or filtering initiated by user actions).
\end{enumerate}
The knowledge base serves as a detailed, machine-readable catalog of VA system architectures, grounded in published research.
The consistent structure across all system entries facilitates systematic analysis and comparison.

\subsection{A visual interface for knowledge base exploration}
\label{sec:interface}
\begin{figure}[t]
\centering
\includegraphics[width=0.9\linewidth]{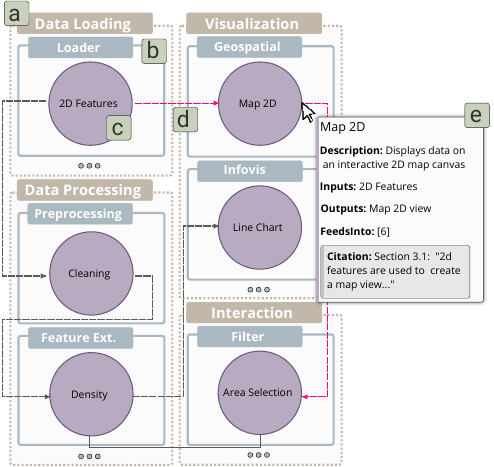}
\caption{%
\projectname's visual interface displaying a summarized system's blueprint.
(a) High-level blocks (e.g., \protect \rectanglelegend{highlevelbgcolor}{Data Loading}) spatially organize (b) intermediate functional blocks (e.g., \protect \rectanglelegend{intermediatebgcolor}{Loader}). 
(c) Nodes within these represent specific granular blocks (e.g., \protect \rectanglelegend{granularbgcolor}{2D Features}). 
Edges depict dependencies: \protect\inlinearrow[0.8em][-0.1em]{figs/dashed} indicates data flowing, while \protect\inlinearrow[0.2em][0.25em]{figs/straight} represents interactions.
(d) Hovering over a node highlights connected edges \protect\inlinearrow[0.8em][-0.1em]{figs/dashed_pink}; (e) clicking a node reveals its detailed information derived from the JSON blueprint.
}
\label{fig:interface}
\end{figure}

To aid in the manual construction, validation, and exploration of the knowledge base, we developed an interactive web-based interface. 
This interface takes the JSON file for a single VA system as input and renders its architecture visually, as exemplified in Figure \ref{fig:interface}.
The interface employs a node-link diagram metaphor, spatially organizing components according to the \projectname\ structure:
\begin{itemize}[noitemsep, topsep=0pt, leftmargin=*]
    \item \textbf{Spatial Grouping:} The visual layout uses spatial organization to reflect the \projectname's hierarchy. Nodes representing granular components are nested within intermediate category boxes, which are inside high-level stage areas. For example, a \rectanglelegend{granularbgcolor}{Map 2D} node placed inside a \rectanglelegend{intermediatebgcolor}{Geospatial} box within the larger \rectanglelegend{highlevelbgcolor}{Visualization} region.
    \item \textbf{Links:} Edges are drawn between nodes to represent the documented data and interaction dependencies, illustrating the flow and control structure of the system.
\end{itemize}

This visual representation proved invaluable during the manual curation phase (Section~\ref{sec:manual}), allowing us to maintain a clear overview of a system's components and their interconnections while extracting information.
It also served as a crucial tool during the knowledge base assessment (Section~\ref{sec:feedback}), providing system builders with an accessible and interpretable view of their own systems as captured in the knowledge base.
For researchers and developers using the knowledge base, the interface offers a way to visually explore and understand the specific architecture of individual VA systems documented within the \projectname.
Both the \projectname\ knowledge base and the visualization interface are publicly available as open-source resources.

\section{Analysis and Application of the Knowledge Base}

In this section, we evaluate the knowledge base through high-level statistics and targeted analyses.
Section~\ref{sec:overview} provides an overview, including component and dependency distributions.
Section~\ref{sec:topology} analyzes the network topology of component graphs to reveal structural patterns.
Section~\ref{sec:temporal} examines how VA system structure and complexity have changed over time.
Finally, Section~\ref{sec:applications} demonstrates applications of the knowledge base by dissecting influential and structurally complex VA systems.


\begin{figure}[t]
\centering
\includegraphics[width=0.85\linewidth]{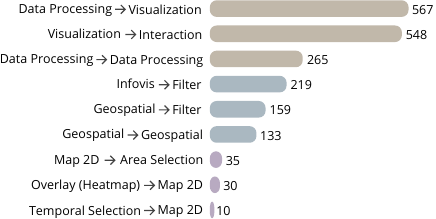}
\caption{Most frequent component dependencies (edges) identified across the knowledge base, categorized by hierarchy level: \protect\circl{highlevelbgcolor}{}~High, \protect\circl{intermediatebgcolor}{}~Intermediate, \protect\circl{granularbgcolor}{}~Granular. Frequency counts are shown on the right.
}
\vspace{-0.5cm}
\label{fig:barchart}
\end{figure}

\subsection{A snapshot of VA systems}
\label{sec:overview}

The final knowledge base consists of structured specifications for 101 VA systems. Each system is represented as a hierarchical dataflow model composed of components, operations, and dependencies, following the formal schema introduced earlier.
In total, the knowledge base contains 403 high-level blocks, 756 intermediate-level blocks, and 1,434 granular-level blocks. 2,448 edges were extracted, with 78.4\% data-driven and 21.6\% interaction ones.
The number of components per system ranged from 9 to 39, with an average of 25 components per system. Considering the abstraction levels, there was an average of 4 high-level components per system, 7 intermediate-level components, and 14 granular-level components.
The number of edges per system ranged from 8 to 62, with an average of 24 edges per system (19 \textit{data} and 5 \textit{interaction} dependencies).

Across all systems, the most common intermediate-level components include \rectanglelegend{intermediatebgcolor}{Loader} (100\%), \rectanglelegend{intermediatebgcolor}{Geospatial} processing modules (99\%), and \rectanglelegend{intermediatebgcolor}{Filter} mechanisms (95\%). Nearly all systems also include at least one \rectanglelegend{intermediatebgcolor}{Infovis} element (94\%). Among analytical components, \rectanglelegend{intermediatebgcolor}{Clustering} appears most frequently (25\%).
At the granular level, the most frequently used components are \rectanglelegend{granularbgcolor}{Map 2D} views, appearing in 80\% of the systems. These are followed by interactive selection mechanisms, such as \rectanglelegend{granularbgcolor}{Area Selection} (49\%) and \rectanglelegend{granularbgcolor}{Temporal Selection} (38\%), as well as common visualization designs such as \rectanglelegend{granularbgcolor}{Heatmap} overlays (36\%) and \rectanglelegend{granularbgcolor}{Line charts} (35\%).

We also analyzed which systems have the highest complexity in terms of their internal structure. Based on the number of granular components, the most complex systems are VAUD (27 blocks) \cite{chen_vaud_2018}, MobiSeg (22) \cite{wu_mobiseg_2017}, and Urban Rhapsody (20) \cite{rulf_urban_2022}.
In terms of edge complexity, the most intricate systems are TPFlow (62 total dependencies \cite{tpflow_liu_2019}; 46 \textit{data}, 15 \textit{interaction}), SemanticTraj (58 total; 52 \textit{data}, 6 \textit{interaction}) \cite{al-dohuki_semantictraj_2017}, and HORA 3D (52 total; 33 \textit{data}, 19 \textit{interaction}) \cite{rauer-zechmeister_hora3d_2024}. 

\subsection{Component topology and interconnections}
\label{sec:topology}

To understand how VA systems are architected, we analyzed the underlying component network structures, including component type distribution and connection topology.
Across all systems, \rectanglelegend{highlevelbgcolor}{Visualization} components were most prevalent (39.7\%), followed by \rectanglelegend{highlevelbgcolor}{Data Processing} (25.5\%), interaction mechanisms (21.6\%), and \rectanglelegend{highlevelbgcolor}{Data Loading} modules (13.2\%).
We also examined dependency edges.
Figure \ref{fig:barchart} shows the most frequent dependencies identified across the knowledge base, detailing common connection patterns at the high, intermediate, and granular levels.
At the high level, the dominant flows are from \rectanglelegend{highlevelbgcolor}{Data Processing} $\rightarrow$ \rectanglelegend{highlevelbgcolor}{Visualization} (567 edges) and from \rectanglelegend{highlevelbgcolor}{Visualization} $\rightarrow$ \rectanglelegend{highlevelbgcolor}{Interaction} (548 edges).
As shown in the figure, at the intermediate level, \rectanglelegend{intermediatebgcolor}{Filter} components frequently mediate interactions with both \rectanglelegend{intermediatebgcolor}{Infovis} and \rectanglelegend{intermediatebgcolor}{Geospatial} components.
For granular connections, \rectanglelegend{granularbgcolor}{Map 2D} views frequently connect to \rectanglelegend{granularbgcolor}{Area Selection} (35 total edges).

We also identified central components -- those with the highest total degree (sum of in-degree and out-degree). 
At the high level, \rectanglelegend{highlevelbgcolor}{Visualization} components function as the most connected hubs (1,828 total degree). \rectanglelegend{highlevelbgcolor}{Data Processing} modules exhibit balanced connectivity (746 incoming, 846 outgoing connections total). 
At the intermediate level, \rectanglelegend{intermediatebgcolor}{Geospatial} (927) and \rectanglelegend{intermediatebgcolor}{Infovis} (898) are the most connected, while \rectanglelegend{intermediatebgcolor}{Filter} components are the most interaction-heavy (778). 
At the granular level, the \rectanglelegend{granularbgcolor}{Map 2D} component is the most central node (357), followed by \rectanglelegend{granularbgcolor}{Area Selection} (167) and \rectanglelegend{granularbgcolor}{3D Scene} (123).

\subsection{Temporal evolution of VA systems}
\label{sec:temporal}

\begin{figure}[t]
\centering
\includegraphics[width=1\linewidth]{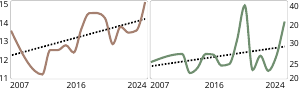}
\caption{Temporal evolution of the average number of blocks and dependencies in VA systems. \protect\circl{blockslinecolor}{}~Mean number of blocks (left) and \protect\circl{edgeslinecolor}{}~Mean number of edges (right). \protect\circl{black}{} Dotted lines indicate the overall upward trend.}
\vspace{-0.5cm}
\label{fig:linechart}
\end{figure}

To understand how VA systems evolved, we analyzed trends in complexity, interactivity, visualization practices, and component diversity across the 101 systems in our knowledge base, spanning from 2007 to 2024.
Figure~\ref{fig:linechart} shows yearly averages for these trends.

\myparagraph{Growing structural complexity}
VA systems have grown more complex over the past two decades, in terms of both size and connectivity.
The \emph{average number of granular components} per system has increased from 13.5 in 2007 to 15.8 in 2024.
The \emph{average number of edges} -- representing both data and interaction dependencies -- has shown an even steeper rise, growing from 21.8 in 2007 to 30.4 in 2024, a 38.4\% increase.
Data from 2019 shows the highest connectivity within our corpus, with an average of 35.9 dependencies per system and an edge-to-node ratio of 2.4 (average is 0.9).

\myparagraph{Breadth over quantity}
We did not observe an upward or downward trend in the proportion of interactive components over time.
Across all years, \rectanglelegend{granularbgcolor}{Area Selection} and \rectanglelegend{granularbgcolor}{Temporal Selection} emerged as the most consistently used techniques.

\myparagraph{Evolving visualization and analytical practices}
2D map visualizations have remained central, appearing in 80\% of systems analyzed.
Recent systems show a trend towards more complex maps; average map overlay components (e.g., \rectanglelegend{granularbgcolor}{Heatmap}) rose from 0.8 per system before 2017 to 1.2 after.
For example, EpiMob \cite{yang_epimob_2023} layers heatmaps, scatter plots, and routes on maps.
Use of \rectanglelegend{granularbgcolor}{3D Scene} has grown notably; in 2024, 4 of 9 systems used 3D visualizations~\cite{boorboor_submerse_2024,lyu_if-city_2024,rauer-zechmeister_hora3d_2024,wagner_reimaging_2024}.

\myparagraph{Shifts in component diversity}
Component diversity within major categories also increased over time.
Comparing the 2017-2024 period (59 systems) to 2007-2016 (42 systems), the number of unique granular component types identified grew substantially: \rectanglelegend{highlevelbgcolor}{Data Processing} (\textasciitilde+240\%), \rectanglelegend{highlevelbgcolor}{Visualization} (\textasciitilde+130\%), \rectanglelegend{highlevelbgcolor}{Data Loading} (\textasciitilde+260\%), and \rectanglelegend{highlevelbgcolor}{Interaction} (\textasciitilde+160\%)
Despite these shifts, visualization has remained the dominant class of component throughout the period, consistently comprising between 35\% and 45\% of system architecture.

\subsection{Applications of the knowledge base}
\label{sec:applications}

Beyond structural analyses and statistical insights, our knowledge base serves as a lens for examining individual VA systems.
In this section, we show how \projectname\ can be used to revisit systems.
Rather than a replacement for the papers or their narrative contributions, \projectname\ provides a complementary structural summary, a compact way to understand a system's architecture and to compare it with others.

\subsubsection{Example 1: Highly influential VA}
\begin{figure*}[h!]
  \centering
  \includegraphics[width=0.9\textwidth]{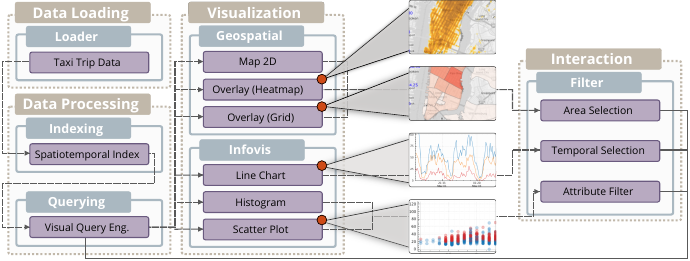}
  \caption{%
  Taxivis's blueprint.
  Loading \protect\rectanglelegend{granularbgcolor}{Taxi Trip Data}, \protect\rectanglelegend{granularbgcolor}{Spatiotemporal Index} and the \protect\rectanglelegend{granularbgcolor}{Visual Query Engine}.
  Results feed into \protect\rectanglelegend{intermediatebgcolor}{Geospatial} and \protect\rectanglelegend{intermediatebgcolor}{Infovis} components, with visualizations for \protect\rectanglelegend{granularbgcolor}{Heatmap} overlay and \protect\rectanglelegend{granularbgcolor}{Line Chart}.
  \protect\rectanglelegend{intermediatebgcolor}{Filter} interactions loop back to the query engine.}
  \label{fig:taxivis}
  \vspace{-0.5cm}
\end{figure*}

We demonstrate the knowledge base's application using TaxiVis \cite{ferreira_visual_2013} as our first example.
TaxiVis is a highly influential system (>700 citations) for exploring NYC taxi data.
We dissect its architecture, pipeline, analytics, visualizations, interactions, and complexity relative to the knowledge base.
Figure \ref{fig:taxivis} shows the system's blueprint.

\myparagraph{System organization} 
TaxiVis' intermediate-level architecture consists of six primary components: \rectanglelegend{intermediatebgcolor}{Loader} for \rectanglelegend{highlevelbgcolor}{Data Loading}; \rectanglelegend{intermediatebgcolor}{Indexing}, and \rectanglelegend{intermediatebgcolor}{Querying} for \rectanglelegend{highlevelbgcolor}{Data Processing}; \rectanglelegend{intermediatebgcolor}{Geospatial} and \rectanglelegend{intermediatebgcolor}{Infovis} for \rectanglelegend{highlevelbgcolor}{Visualization}; and \rectanglelegend{intermediatebgcolor}{Filter} for \rectanglelegend{highlevelbgcolor}{Interaction}.
At the granular level, these components are: \rectanglelegend{granularbgcolor}{Taxi Trip Data} input, a \rectanglelegend{granularbgcolor}{Spatiotemporal Index}, a \rectanglelegend{granularbgcolor}{Visual Query Engine}, visualization techniques including \rectanglelegend{granularbgcolor}{Map 2D}, \rectanglelegend{granularbgcolor}{Heatmap}, \rectanglelegend{granularbgcolor}{Grid}, \rectanglelegend{granularbgcolor}{Line Chart}, \rectanglelegend{granularbgcolor}{Histogram}, and \rectanglelegend{granularbgcolor}{Scatter Plot}, and interaction mechanisms comprising \rectanglelegend{granularbgcolor}{Area Selection}, \rectanglelegend{granularbgcolor}{Temporal Selection}, and \rectanglelegend{granularbgcolor}{Attribute Selection}.

\myparagraph{Comparison to the knowledge-base}
TaxiVis features 6 intermediate-level blocks and 12 granular blocks (15\% fewer than the knowledge base average of 14.2).
The system comprises 14 dataflow edges and 3 interaction edges, 29.8\% fewer than the knowledge base average of 24.2.

\myparagraph{Data processing pipeline}
TaxiVis processes data using a specialized k-d tree \rectanglelegend{granularbgcolor}{Spatiotemporal Index} for interactive querying. The \rectanglelegend{granularbgcolor}{Visual Query Engine} translates user interactions into structured queries. 
Raw NYC taxi dataflows: \rectanglelegend{granularbgcolor}{Taxi Trip Data} $\rightarrow$ \rectanglelegend{granularbgcolor}{Spatiotemporal Index} $\rightarrow$ \rectanglelegend{granularbgcolor}{Visual Query Engine} $\rightarrow$ \rectanglelegend{highlevelbgcolor}{Visualization} components.

\myparagraph{Visualization components}
TaxiVis uses six visualizations: a primary \rectanglelegend{granularbgcolor}{Map 2D} with a \rectanglelegend{granularbgcolor}{Heatmap} overlay for density and a \rectanglelegend{granularbgcolor}{Grid} overlay for regional statistics;
a \rectanglelegend{granularbgcolor}{Line Chart} for temporal patterns; a \rectanglelegend{granularbgcolor}{Histogram} for attribute distributions;
and a \rectanglelegend{granularbgcolor}{Scatter Plot} for attribute relationships.

\myparagraph{Analytics framework}
TaxiVis integrates analytics via its \rectanglelegend{granularbgcolor}{Visual Query Engine}, linking data selection to visualization.
Views propagate selections across spatial, temporal, and attribute dimensions.
Visualizations support direct manipulation, creating an exploration loop where selections feed back into the query engine (Figure \ref{fig:taxivis}).

\subsubsection{Example 2: Most complex VA by block count}
In this section, we analyze VAUD \cite{chen_vaud_2018}. 
The system stands out since it presents the highest number of granular blocks extracted.
Our analysis then focuses on the distribution of components relative to dataset averages, the observed flow patterns, factors contributing to the high number of blocks, and the mechanisms employed to manage complexity.

\myparagraph{System scale}
VAUD has the most granular blocks (27 vs. 14.2 average) and 8 intermediate blocks, totaling 35 (vs. 25.7 average).
Its 41 edges (37 \textit{data}, 4 \textit{interaction}) exceed the knowledge base average of 24.2. Granular blocks are distributed across \rectanglelegend{highlevelbgcolor}{Data Loading} (8), \rectanglelegend{highlevelbgcolor}{Visualization} (8), \rectanglelegend{highlevelbgcolor}{Data Processing} (6), and \rectanglelegend{highlevelbgcolor}{Interaction} (5).

\begin{figure}[h]
\centering
\includegraphics[width=0.9\linewidth]{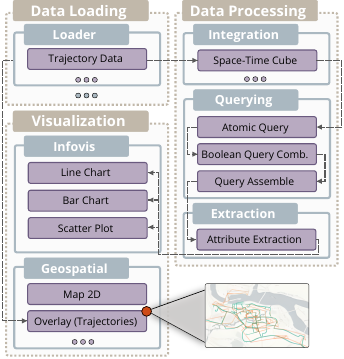}
\caption{%
VAUD's simplified blueprint view tracing the path for loaded \protect\rectanglelegend{granularbgcolor}{Trajectory Data}.
While this data proceeds through \protect\rectanglelegend{intermediatebgcolor}{Integration} (\protect\rectanglelegend{granularbgcolor}{Space-Time Cube}), \protect\rectanglelegend{intermediatebgcolor}{Querying}, and \protect\rectanglelegend{intermediatebgcolor}{Extraction} for integrated analysis with other data types, it also directly feeds the \protect\rectanglelegend{granularbgcolor}{Trajectories} overlay to be displayed on the map visualization.
}
\vspace{-0.7cm}
\label{fig:vaud}
\end{figure}

\myparagraph{Dataflow patterns}
The system integrates diverse urban data sources (e.g., POIs, taxi trajectories, social networks) through a unified query model, with data flowing from multiple loaders into shared visualization components.
A representative flow in VAUD begins, for instance, by loading raw \rectanglelegend{granularbgcolor}{Trajectory Data} in conjunction with other urban data. Sequentially, the data is normalized in the \rectanglelegend{granularbgcolor}{Space-Time Cube}, processed by the \rectanglelegend{granularbgcolor}{Atomic Query}, refined by \rectanglelegend{granularbgcolor}{Boolean Query Combination} and \rectanglelegend{granularbgcolor}{Query Assemble}, attributes extracted (\rectanglelegend{granularbgcolor}{Attribute Extraction}), and finally displayed on the \rectanglelegend{intermediatebgcolor}{Infovis} components (Figure \ref{fig:vaud}).

\myparagraph{High count source}
VAUD's high block count stems from: (1) integrating diverse data sources (increasing \rectanglelegend{intermediatebgcolor}{Loader} components);
(2) a layered query model with 3 dedicated blocks (\rectanglelegend{granularbgcolor}{Atomic Query}, \rectanglelegend{granularbgcolor}{Boolean Query Combination}, \rectanglelegend{granularbgcolor}{Query Assemble}), followed by a processing model (\rectanglelegend{granularbgcolor}{Attribute Extraction}) that has to feed an above-average number of interactive visualizations.

\myparagraph{Complexity management}
VAUD manages complexity via a canonical space-time representation unifying heterogeneous data for cross-domain analysis.
This unified approach is complemented by a layered query model that decomposes analytical tasks into discrete steps (atomic queries and attribute extractions).
The system also separates visualization concerns with dedicated components for different data types (e.g., dedicated overlay to deal with \rectanglelegend{granularbgcolor}{Trajectories}, specific glyphs to handle \rectanglelegend{granularbgcolor}{POI Data} visualization).

\subsubsection{Example 3: Most complex in terms of number of edges}
In this section, we focus on TPFlow \cite{tpflow_liu_2019}, a system whose defining characteristic is its high degree of interconnection among components.
Our analysis centers on the number and distribution of edges relative to dataset norms, the system’s reliance on interaction-driven hub nodes, and the mechanisms that support highly connected VA workflows.

\myparagraph{System connectivity} 
 TPFlow features 18 granular components but the highest edge count (62: 47 \textit{data}, 15 \textit{interaction}) in the knowledge base.
 Its graph density (\textasciitilde0.2) is considerably above the average (0.1), indicating high connectivity, with many nodes forming tightly knit clusters, presenting a clustering coefficient of \textasciitilde0.5 (average is \textasciitilde0.2).

\myparagraph{Key hubs}
High-degree hubs include interaction components (e.g., \rectanglelegend{granularbgcolor}{Brushing}, \rectanglelegend{granularbgcolor}{Linking}, \rectanglelegend{granularbgcolor}{Superposition}, and \rectanglelegend{granularbgcolor}{Juxtaposition}) coordinating input from multiple visual sources (e.g., \rectanglelegend{granularbgcolor}{Line Chart}, \rectanglelegend{granularbgcolor}{Bar Chart}, \rectanglelegend{granularbgcolor}{Map 2D}) and distributing updates.
The system’s tensor decomposition blocks (including \rectanglelegend{granularbgcolor}{Piecewise Decomposition} and \rectanglelegend{granularbgcolor}{Rank-One Approximation}) function as key data processing hubs that feed visual components.

\myparagraph{Dependency-driven analytics} 
TPFlow employs progressive decomposition to structure spatiotemporal analysis.
Partitions and derived patterns are visualized using comparison techniques (e.g., \rectanglelegend{granularbgcolor}{Superposition}, \rectanglelegend{granularbgcolor}{Juxtaposition}). 
Crucially, feedback loops enable dependency-driven refinement: user interactions (e.g., \rectanglelegend{granularbgcolor}{Brushing}) in any linked view can trigger re-decomposition (\rectanglelegend{granularbgcolor}{Piecewise Decomposition}), propagating changes and allowing iterative focusing on specific phenomena.

\myparagraph{Managing complexity through feedback loops} 
The system supports hierarchical exploration via interactive partition refinement (\rectanglelegend{granularbgcolor}{Tree Split}).
User choices (e.g., splitting, re-labeling) branch the analysis, tracked visually by a provenance tree.
Outputs are simultaneously channeled to diverse, coordinated visualizations (e.g., charts, maps), allowing multifaceted comparison and iterative refinement based on user decisions.

\section{Evaluation}
\label{sec:feedback}

To evaluate the VA-Blueprint knowledge base and assess its accuracy, completeness, and usefulness, we conducted a series of interviews with system builders who developed VA systems represented in our knowledge base, \highlight{as well as a quantitative analysis of annotation metrics.}
%

\subsection{Experts' feedback}

\myparagraph{Interview methodology} We interviewed five experts (hereafter referred to as \rectangle{rectanglecolor}{E1-5}) who co-authored nine different VA systems in our knowledge base.
Before the interviews, participants received their extracted system representations and access to our interactive interface to explore the structured components.
Interviews followed a semi-structured format covering (1) accuracy and completeness, (2) effectiveness of the hierarchical representation, and (3) potential usefulness.
Each session lasted 40–60 minutes, with participants providing detailed feedback based on their prior exploration.

\myparagraph{Knowledge base extraction accuracy} 
The overall assessment of blocks and flows extraction accuracy was positive, with researchers confirming that the extracted building blocks captured the essence of their systems.
\rectangle{rectanglecolor}{E1} noted that the extracted components were ``99\% ok'', while \rectangle{rectanglecolor}{E2} described the extraction as ``quite faithful'' to the actual systems. 
\rectangle{rectanglecolor}{E5} was particularly impressed that the approach could extract not just systems' components but also their relationships: ``I think both represent the system globally very well... I found this quite impressive.''

However, participants identified missing components. 
Across nine evaluated systems, interviewees found 7 missing granular blocks and 5 missing edges.
For instance, \rectangle{rectanglecolor}{E3} noted a missing data source crucial to system functionality.
Similarly, \rectangle{rectanglecolor}{E5} observed that one system was shown using only image data, while it actually handled multiple human activity data sources.
\rectangle{rectanglecolor}{E1} highlighted a missing processing component important for retrieval, and \rectangle{rectanglecolor}{E4} identified a missing edge for a feedback loop from an interaction block to a processing block.

\myparagraph{Hierarchical representation}
The three-level hierarchical representation received positive validation from all interviewees. 
\rectangle{rectanglecolor}{E3} affirmed the approach's structural coherence for system component organization, noting that the multi-level abstraction well represents the architectural relationships between components. 
\rectangle{rectanglecolor}{E5} emphasized its cognitive benefits in facilitating system comprehension through categorization.
Participants also offered constructive suggestions to enhance the representation.
\rectangle{rectanglecolor}{E2} proposed extending the hierarchical model to encompass dataflow relationships at multiple abstraction levels: ``Since you're proposing a hierarchical representation of the system, I think it makes sense to hierarchize everything, not just the components.''
Also, \rectangle{rectanglecolor}{E1} and \rectangle{rectanglecolor}{E2} suggested that explicitly representing preprocessing operations and on-the-fly operations would be valuable.
\rectangle{rectanglecolor}{E2} noted: ``distinguishing between operations that happen before the system starts versus those that happen during user interaction is important for evaluating a system's technical capabilities and performance claims.''

\myparagraph{Knowledge share value} 
Beyond applications to development, interviewees noted \projectname's usefulness in multiple applications. 
\rectangle{rectanglecolor}{E1} identified two benefits: ``understanding system evolution'' through versioning similar to GitHub, and ``pruning the system'' by identifying unnecessary components. 
\rectangle{rectanglecolor}{E2} emphasized its value as ``a reference tool'' for understanding how similar systems were developed, and suggested its utility for meta-analysis: ``You could use this to do broader studies of what's being implemented and how it's being implemented... could you detect changes? Are people implementing things the same way they did ten or twenty years ago?''
\rectangle{rectanglecolor}{E3} highlighted VA-Blueprint's value for team collaboration (e.g., design sharing), idea formalization during pre-implementation design phases, and educational purposes.
\rectangle{rectanglecolor}{E4} stressed VA-Blueprint's benefits for planning and implementation, noting how it provides a system overview that facilitates step-by-step task organization and development workflow structuring.
\rectangle{rectanglecolor}{E5} highlighted how the knowledge base could help researchers learn from existing implementations, noting it would be valuable for ``observing other possibilities'' in system design by comparing different systems.

\begin{table}[t!]
\centering
\caption{\highlight{Comparison of manual and LLM annotation statistics.}}
\vspace{-0.25cm}
\begin{tabular}{@{} l c c c @{}}
\toprule
\textbf{Metric} & \textbf{Manual} & \textbf{LLM} & \textbf{Difference} \\
\midrule
Edges & 18.8 $\pm$ 6.77 & 16.65 $\pm$ 6.98 & 6.45 $\pm$ 4.20 \\
Intermediate blocks & 7.65 $\pm$ 1.76 & 6.9 $\pm$ 1.37 & 1.35 $\pm$ 1.23 \\
Granular blocks & 13.6 $\pm$ 3.28 & 11.3 $\pm$ 2.77 & 3.30 $\pm$ 2.25 \\
\bottomrule
\end{tabular}
\vspace{-0.5cm}
\label{table:comparison}
\end{table}

\subsection{Quantitative evaluation}

\highlight{To quantitatively assess the VA-Blueprint knowledge base, we evaluated 20 urban VA papers not included in our initial dataset, chosen to represent diverse domains, groups, and data types. Each paper was manually annotated to create a ground truth, then compared with LLM-generated annotations.}
\highlight{We measured three structural metrics: total edges, intermediate blocks, and granular blocks. Table~\ref{table:comparison} summarizes the results, showing the mean and standard deviation, along with the mean absolute difference across the 20 papers.}

\highlight{On average, manual annotation recorded about 13.6 granular blocks per paper, whereas the LLM identified 11.3. Manual analysis found 13.4 distinct granular block names per paper, compared to 11.25 by the LLM. In terms of system connections, manual analysis averaged 18.8 edges per paper, with 16.65 edges by the LLM.}
\highlight{Performance varied by component: the LLM under-extracted granular blocks by 22\% in Data Processing, 15\% in Visualization, 12\% in Data Loading, and 9\% in Interaction. While the LLM consistently identified common blocks such as \emph{Map 2D} and standard chart types (e.g., \emph{Bar Chart}, \emph{Line Chart}), semantic and structural differences emerged. In two cases, the LLM aggregated distinct functional steps into a single block. For example, in the manual annotation of one system, \emph{Incremental Beneficial Scores} and \emph{Greedy Network Expansion} appeared as separate sequential blocks, while the LLM combined these into a single unit.}
%
%
\highlight{We found an 86\% match rate between manual and LLM block labels. Despite differences, the LLM performed well in extracting block justifications: of 226 granular blocks identified, only two had incorrect or missing citations.}

%

\section{Discussion}

A key contribution of this work is the formalization and operationalization of VA system components into a structured, queryable, and extensible knowledge base using LLMs.
This structure provides a few benefits.
First, it offers a blueprint for system design. By analyzing recurring component patterns, dependencies, and architectures across 101 systems, researchers and practitioners can better understand what design elements are common and how they interconnect.
Second, it supports documentation of systems. The specification offers a formal way to describe VA systems -- beyond text and screenshots -- by capturing structure, logic, and dependencies. This can enable, in future work, a more precise comparison between systems.
Third, the knowledge base can further support tool development. The taxonomy of components can serve as a foundation for future system-building platforms, such as low-code authoring systems that leverage component reuse.

\myparagraph{Balancing automation and expert refinement}
\highlight{The quantitative analysis reveals that while the LLM demonstrates promising capabilities in extracting structured representations of VA systems, there are limitations that reinforce the value of human oversight. Its tendency to aggregate distinct functional steps into single blocks and under-extract connections and processing components points to challenges in capturing the nuanced structure of complex VA systems. However, the efficiency gains offered by the LLM, which reduced annotation time from over 18 hours to under two hours, highlight its potential as a tool for scaling knowledge base construction.}

\myparagraph{Limitations of the current categorization}
Our methodology suffers from semantic ambiguity; components such as ``filter'' or ``aggregate'' appear in different contexts and may serve different roles depending on system goals. Another problem is with respect to component granularity, i.e., what one system considers an atomic operation may be decomposed into multiple components in another.
While we mitigated this through manual revisions, inconsistencies may persist.

\myparagraph{Limitations of the LLM-assisted extraction}
While LLMs proved effective for scaling the knowledge base, their use introduces limitations. The key challenge is navigating the \emph{abstraction spectrum}, i.e., deciding at what level to define a component or operation. Sometimes, the model may overgeneralize, inferring high-level components only loosely described in the paper. In other cases, it struggles with granularity, conflating conceptually distinct operations. For instance, an LLM might identify a component simply as ``clustering'', even though the paper describes two distinct operations: ``spatial clustering'' and ``temporal clustering.'' Without clear contextual grounding or human inspection, the model may either collapse them into one overly general component or separate them.
This ambiguity is particularly evident when system functionality is described in the text or through usage scenarios rather than clear modular descriptions.
Finally, our methodology is still semi-automated, with each LLM-generated specification requiring human oversight, including correction and validation.

\myparagraph{Generalizability and domain specificity}
\highlight{Our schema, with its multi-level component structure and defined relationships, is inherently general and designed for broad applicability across VA domains. Extending our work to new domains wouldn't alter this fundamental schema; rather, our general methodology, combining human-in-the-loop expert coding with LLM-assisted extraction, can be directly applicable to new fields. This process would instantiate the same schema with domain-specific content, creating knowledge bases tailored to those areas. Thus, \projectname\ serves as both a foundational structure and a reusable methodology for systematically understanding VA systems.}


\section{Conclusions and Future Work}

In this paper, we presented a methodology for uncovering and structuring building blocks of VA systems through a combination of manual coding and LLM-assisted extraction.
By analyzing 101 papers, we created \projectname, a multi-level knowledge base that captures the components, operations, and dependencies.
Our findings reveal common design patterns, increasing system complexity over time, and evolving practices.
We also identified key limitations, especially when dealing with the granularity and abstraction of components.
\highlight{Importantly, however, our results suggest that while LLMs can provide a strong baseline for system representation, expert refinement like the one used on VA-Blueprint is essential to ensure accuracy, completeness, and semantic fidelity.}

Looking ahead, our future work will explore several directions.
First, we aim to expand the corpus beyond urban-focused systems to assess the generalizability of our schema across domains.
Second, while our LLM-assisted extraction currently relies on textual content (e.g., captions, references), we aim to enhance this pipeline by incorporating a multimodal approach to include images.
%
%
Third, we will explore leveraging models with larger context windows.
\highlight{Fourth, we plan to integrate additional data sources beyond papers, like manuals and source code, to enhance the knowledge base's granularity and accuracy.}
%
%
Ultimately, we built this knowledge base not as a static artifact, but as a living resource that can evolve with the field -- supporting reflection and reuse in VA system design. \highlight{To maximize its utility, we have made the knowledge base open source, allowing researchers and practitioners to extend, refine, and apply it in their own VA system design efforts.}

\section*{Acknowledgments}
We thank the reviewers for their constructive feedback.
This work was supported by the U.S. National Science Foundation (Awards \#2320261, \#2330565, \#2411223) and conducted as part of the Open-Source Cyberinfrastructure for Urban Computing (OSCUR) project.

\bibliographystyle{abbrv-doi-hyperref}

\bibliography{references}


\end{document}